%% file: 0_Main.tex
\documentclass[manuscript]{acmart}

\AtBeginDocument{%
  }

\acmConference[Academic Mindtrek '26]{International Academic Mindtrek Conference 2026}{October 27-30, 2026}{Tampere, Finland}

\setcopyright{none}

\usepackage{cleveref}
\usepackage{enumitem}
\usepackage[utf8]{inputenc}
\usepackage{amsmath}
\usepackage[ruled,vlined]{algorithm2e}
\usepackage{longtable}
\usepackage{array}
\usepackage{multibib}
\newcites{Ludo}{Ludography}

\begin{document}

\title{Become the Beast: Exploring Human-Quadruped Locomotion for Exergames}

\author{Shamit Ahmed}
\orcid{0009-0002-7462-9760}
\email{shamit.ahmed@aalto.fi}
\affiliation{%
  \institution{Aalto University}
  \city{Espoo}
  \country{Finland}}

\author{Prabhav Bhatnagar}
\orcid{0000-0001-7196-5254}
\email{prabhav.bhatnagar@aalto.fi}
\affiliation{%
  \institution{Aalto University}
  \city{Espoo}
  \country{Finland}}

\author{Perttu Hämäläinen}
\orcid{0000-0001-7764-3459}
\email{perttu.hamalainen@aalto.fi}
\affiliation{%
  \institution{Aalto University}
  \city{Espoo}
  \country{Finland}}


\begin{abstract}
 
Embodying non-human characters and exercising abdominal muscles are both underexplored in exergames. We address this by describing the design and evaluation of a novel human quadruped locomotion exergame, Become the Beast. In the game, the player lies supine on the ground and moves their arms and legs to control a quadrupedal character (a tiger), similar to common bodyweight abdominal muscle exercises such as the Bicycle Crunch. The motion tracking is computer vision-based, utilizing a Kinect sensor placed above the player, which makes our approach suitable for commercial premises such as indoor activity parks where a system needs to run unattended and without any wearable components. Our system extends embodied interaction beyond traditional bipedal or controller-based systems, demonstrating how natural limb movements can generate responsive and immersive quadrupedal motion within virtual environments. We conducted a user study (N=15) and utilized Reflexive Thematic Analysis (RTA) to evaluate the system's intuitiveness, control, and overall player experience. The findings validate that natural body movements effectively control the avatar while delivering an intense core workout. Notably, gameplay immersion masked physical exertion, allowing rigorous core training to be primarily perceived as play.

\end{abstract}


\begin{teaserfigure}
  \includegraphics[width=\textwidth]{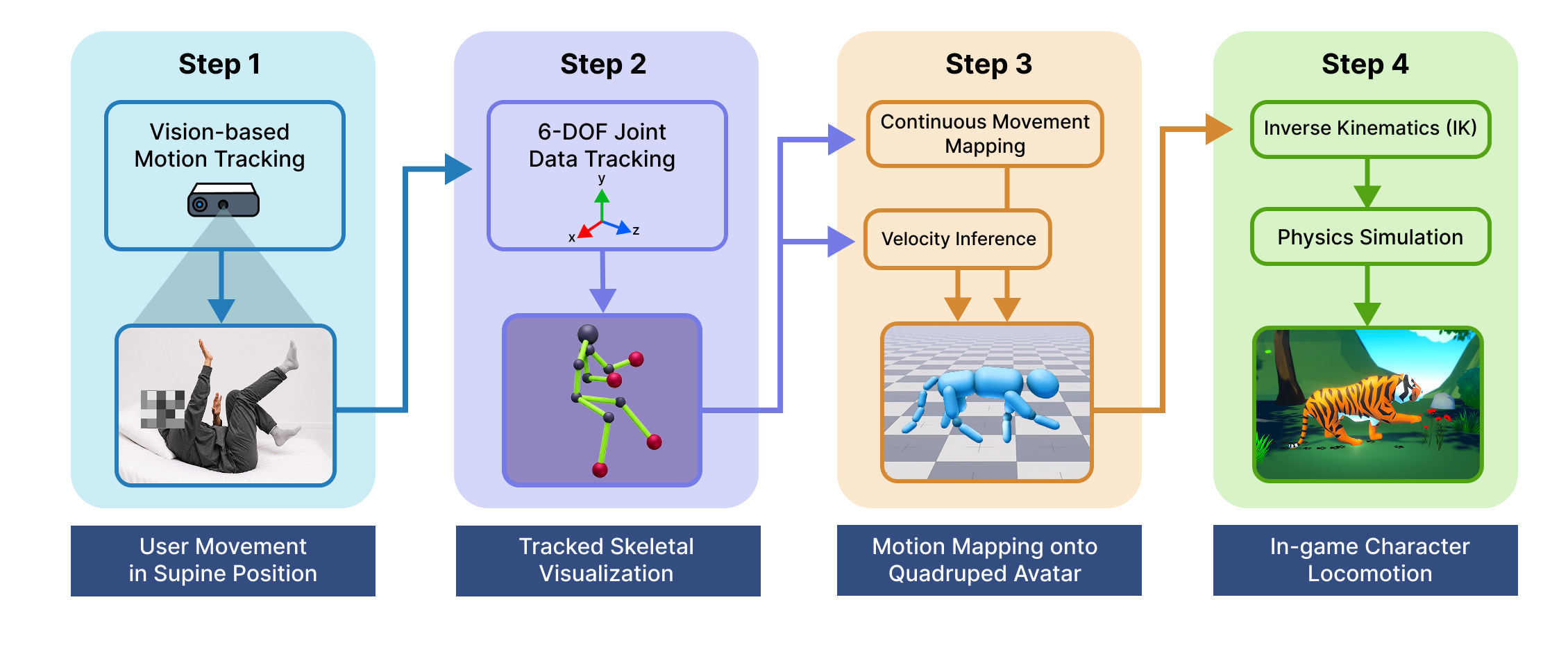}
  \caption{Overview of the proposed human quadruped locomotion system, from left to right: Step 1. User performing movement while in supine position, Step 2. Human skeletal view based on tracked user movement, Step 3. Human motion mapped onto a virtual quadrupedal character, Step 4. Final quadruped locomotion within the game environment. See the supplemental video for more (url: https://youtu.be/IdfNDIQBcL0)}
  \Description{}
  \label{fig:teaser}
\end{teaserfigure}


\maketitle


\input{1_Introduction}
\input{2_RelatedWork}

\input{3_Design}

\input{4_UserStudy}
\input{5_Results}
\input{6_Discussion}

\input{7_Limitations_and_Future_Work}
\input{8_Conclusion}

\bibliographystyle{ACM-Reference-Format}
\bibliography{references}


\clearpage

\appendix
\onecolumn
\input{Appendix} \label{appendix}

\clearpage




\end{document}

%% file: 1_Introduction.tex
\section{Introduction} \label{introduction}

Human motion has long served as a rich foundation for interaction in games and virtual environments. Using body movement as an input modality can offer more intuitive, embodied, and physically engaging experiences compared to traditional controller-based systems \cite{dourish2001action}. Within the field of human-computer interaction (HCI), such embodied interfaces have been studied extensively \cite{Schultze_embodimentreview_2010}, highlighting their potential to foster natural interaction, immersion, and emotional engagement.

Motion-based virtual reality (VR) interaction enhances presence, agency, and realism by aligning user intention with body-driven input \cite{Templeman_walkinplace_1999, usoh_walking_1999}. VR locomotion techniques, however, have their own limitations and can be a source of cybersickness \cite{laviola2000discussion}, reducing usability \cite{coomer_vrlocomotion_2018, Langbehn_vrlocomotion_2018}. Previous work by \citet{reetu_abs_2023} explored supine locomotion in VR and demonstrated its suitability for creating exergames centered on abdominal muscle exercise; however, their approach has two main problems. First, the hardware setup is not accessible, requiring feet trackers that most VR users do not have. Additionally, due to hygiene reasons, a VR setup is not optimal for commercial venues compared to screen-and-camera-based setups. Second, the interaction of \citet{reetu_abs_2023} does not provide meaningful interaction for the hands or allow the user to embody themselves as the avatar using the upper body.

Parallel to this, studies show that physically engaging digital play can promote fitness, social connection, and long-term adherence to exercise routines \cite{mueller_exertion_2003}. This convergence of play, movement, and technology continues to redefine how users experience digital environments, bridging entertainment, well-being, and embodied expression. Furthermore, advances in computer vision (CV) and motion capture technologies have expanded access to full-body interaction. Commodity depth sensors such as the Microsoft Kinect, RGB cameras, and AI-driven pose estimation frameworks now enable real-time tracking of human motion without specialized equipment \cite{shotton_poseRecognition_2011, zheng_DLposeEstimation_2023}.

Locomotion plays a central role in shaping a user’s sense of presence within virtual environments \cite{slater_presence_1999, usoh_walking_1999}. Many existing locomotion approaches rely on hand-held controllers, treadmills, or redirected walking \cite{razzaque_redirected_2005} techniques to simulate movement in virtual spaces. However, designing effective locomotion techniques remains a persistent challenge in HCI, as these systems can be costly, spatially demanding, or cognitively restrictive, often reducing the user’s sense of natural embodiment \cite{nguyen_survey_2021}. This has led to a growing interest in more accessible, full-body alternatives that leverage natural movement as an input modality \cite{Templeman_walkinplace_1999, McMahan_naturalinteraction_2010, Schultze_embodimentreview_2010}.

    

Our motivation is to design a novel locomotion framework that allows users to engage in full-body movement, focusing on addressing the following exploratory research questions: 

\begin{enumerate}
    \item RQ1: What are the player observations and thoughts regarding the overall experience offered by the locomotion technique?
    \item RQ2: How does the locomotion technique feel as a core muscle workout exergame?
    \item RQ3: What design improvements, alternative approaches, and additional features should be prioritized in future iterations?
\end{enumerate}

\vspace{5mm}
\textit{Contribution:} In this paper, we present an HCI artifact contribution \cite{wobborock_contribution_2016} consisting of a locomotion system that translates human motion into quadrupedal character movement in real-time. To demonstrate its utility, we developed an associated exergame prototype designed for exercising abdominal muscles (\Cref{fig:teaser,fig:gameplay}).

Our screen-based setup minimizes cybersickness by eliminating the visual-vestibular conflict typically associated with VR displays. Furthermore, eliminating wearable trackers facilitates quick player switching in social settings and increases commercial viability by enabling unstaffed operation in venues like indoor activity parks.

Our work aligns with ongoing research on exergames, which aims to increase motivation and enjoyment in exercise through interactive and embodied experiences \cite{mueller_exertion_2003, bianchi_engage_2007}. To the best of our knowledge, we are the first to implement vision-based real-time control of virtual quadrupedal locomotion through 4-limb motion input. Beyond its immediate application, our system contributes to embodied interaction and cross-species locomotion design, opening new opportunities in exergames \cite{Naderi_CAI_2018}, physical rehabilitation \cite{gil_ARphysicaltherapy_2021}, and other interactive systems where natural and intuitive movement is essential \cite{McMahan_naturalinteraction_2010}.

%% file: 2_RelatedWork.tex
\begin{figure}
  \centering
  \includegraphics[width = 0.6 \linewidth]{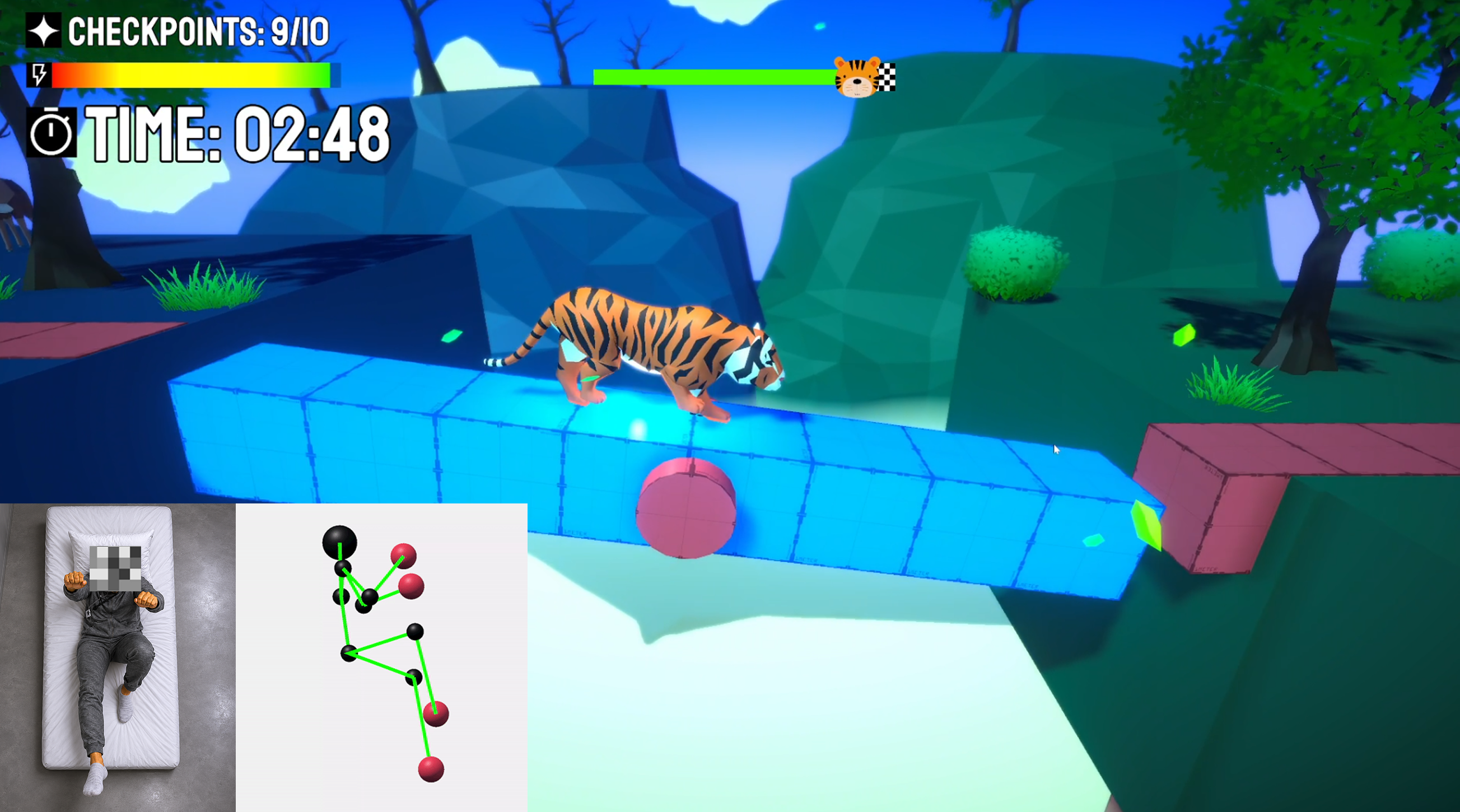}
  \caption{Gameplay screenshot of user playing the game. Note the bottom-left corner displaying the live RGB camera feed alongside the real-time skeletal tracking visualization.}
  \Description{}
  \label{fig:gameplay}
\end{figure}

\section{Related Work}
Below, we review related user-avatar motion mapping and locomotion techniques across fields such as exergames, virtual reality, and computer vision.

\subsection{Human-Quadruped Motion Mapping}

Motion mapping/retargeting aims to transfer motion patterns from one skeletal structure to another while maintaining physical plausibility \cite{Gleicher_retarget_1998}. In the context of bipedal movement targeted to quadrupedal characters, this task becomes particularly challenging due to substantial anatomical and biomechanical differences between bipeds and quadrupeds.

Recent work by \citet{egan_dogcode_2024} presents Dog Code, a deep learning–based framework that maps human motion to quadrupedal movement in VR. Their model demonstrates improved performance for symmetric quadrupedal gaits (walk, trot, pace) compared to traditional inverse kinematics (IK)-based retargeting solutions, but falls short for faster, asymmetric gaits (gallop, canter), likely due to the lack of temporal alignment with fast, symmetric human gaits (jog, run). Moreover, the source motion in Dog Code is bipedal, whereas we study a system where the human is actually mimicking quadrupedal movement in a supine position. Understanding a full-body mapping system like ours can potentially unlock novel abdominal or core workout exercises, which is a highly underexplored area in exergaming.

Virtual animal embodiment, i.e., a user inhabiting a non-humanoid body in a virtual environment, has been explored in multiple studies, especially in VR \cite{Krekhov_VRAnimals_2018, Jiang_HandAvatar_2023}. \citet{Krekhov_VRAnimals_2018} investigated various non-human character body ownership techniques, including a full-body movement technique where users had to kneel and crouch on a yoga mat. Furthermore, the sense of embodiment in non-human avatars is closely linked to visual perspective. \citet{hoppe2022there} demonstrated that the choice between first-person and third-person viewpoints directly affects a user’s sense of body ownership in VR. This is particularly relevant when mapping human movements to a quadrupedal form, as the perspective can either bridge or widen the perceived gap between the player’s physical body and their virtual animal counterpart.

\subsection{Exergames}

Exergames, or "exercise video games," are games that require players to use their bodies to play, turning physical movement into engaging workouts. Exergaming has become an increasingly popular field that involves using immersive technologies to motivate physical activity, such as strength training, balance, aerobic exercise, and flexibility. Multiple studies \cite{bianchi_engage_2007, bianchi_engagement_2013} have also tried to understand the relationship between body movement and player engagement. 

Based on results from 29 studies, \citet{Davis_exergamereview_2024} found that exergames can elicit moderate to vigorous exercise intensity comparable to traditional workouts while significantly lowering the user's perceived exertion and enhancing enjoyment. Consequently, they conclude that exergaming is a viable, engaging substitute for traditional exercise that may improve adherence among sedentary individuals. While recent work by \citet{sebastian_leg_2023} highlights the physiological importance of lower-body engagement, current locomotion paradigms, such as Walking-in-Place (WiP) \cite{usoh_walking_1999}, remain predominantly bipedal. Consequently, non-upright modalities like supine movement are underexplored, representing a significant gap in gamifying core intensive exercise.

Nintendo's Wii Sports \cite{nintendo_wiisports_2006} and Wii Fit \cite{nintendo_wiifit_2007} are widely recognized as the pioneering works that popularized the exergame genre. By introducing intuitive motion controls and the Wii Balance Board, they democratized "exertion interfaces" \cite{mueller_exertion_2003}, proving that gamified physical activity could be accessible, commercially viable, and effective at shifting gaming away from purely sedentary paradigms. Research into non-standing exergames highlights how physical posture significantly influences player experience and exertion. For instance, \citet{gerling2014effects} explored the differences between sitting and standing play, providing a foundation for floor-based exergames and supine interaction models that cater to diverse physical abilities.

\subsection{VR Locomotion Techniques}

Although our design centers on a vision-based, non-VR locomotion system, our approach has been significantly inspired and informed by innovative locomotion techniques across different mediums. Techniques in VR motivated us to study the strengths and drawbacks of physical movement as an input modality. Existing literature reviews frequently compile and compare a range of VR locomotion approaches \cite{nilsson_15years_2018, boletsis_empirical_2019, boletsis_typology_2022}. VR has been well explored due to its novelty and ability to create immersive, embodied human-computer interactions.

Our approach can be classified as a WiP virtual locomotion technique, where the user performs physical motions to move their avatar through the virtual environment while remaining stationary in their physical space \cite{Templeman_walkinplace_1999}. WiP eliminates the need for a large physical space matching the virtual space. Various WiP variants have been proposed over the years. Movement in WiP methods is commonly tracked using wearable motion trackers or omnidirectional treadmill-like devices \cite{darken1997omni}. Although some VR WiP research claims reduced motion sickness when compared to joystick locomotion, WiP in VR is fundamentally problematic as it cannot avoid vection or visual-vestibular conflict, which is a major cause of motion sickness \cite{Chang_2020_VRSickness}. \citet{reetu_abs_2023} also noticed some of their users experiencing motion sickness in their supine locomotion setup. Moreover, using a VR headset while lying down can cause an upturned redirection of the user's cognitive space, resulting in increased visual-vestibular sensory conflict, as investigated by \citet{luo_sensoryconflict_2022}. Motivated by these problems, we implement our game using a screen instead of a VR headset, which avoids the visual-vestibular conflict.

\subsection{Vision-based Locomotion Techniques and Exergames}

Computer vision as an interaction technique can provide natural and human-centered interactions with computers that are robust and accessible. Analysis of human motion through visual information has been a highly active research topic in the HCI community \cite{freeman_CVinteraction_1998, D'Hooge_intel_2001, johanna_shadowboxer_2004, perttu_CVChildrengame_2003}.
At present, CV-based approaches can support multiplayer participation in the same physical space with minimal to no additional resources. According to \citet{peng_2013_multiplayer}, this can result in higher player engagement and enjoyment through communication, collaboration, or even competition against one another.

The idea of physical participation in a virtual environment via vision-based movement tracking began as early as 1969 by \citet{Krueger1983-KRUAR}, where a participant's movements were perceived and responded to by a computer. Their later pioneering work, Video Place \cite{Krueger_videoplace_1985}, explored real-time human-computer interaction, initially implemented as a communication method.

Microsoft Kinect's \cite{kinect} launch in 2010 commercialized and mass-distributed vision-based exergames among the general population. \citet{perttu_martialarts_2005} explored the fitness and entertainment aspects of exergames through their artificial reality martial arts game, Kick Ass Kung-Fu \cite{KickAss}. They also utilized CV tracking to digitally augment traditional exercise equipment, specifically trampolines \cite{perttu_trampoline_2013} and climbing walls \cite{perttu_wall_2016}. \citet{raine2015motiongames} explores how screen-and-vision-based exergames use digital feedback and game mechanics to make physical sports training more motivating and effective for skill development.

Commercial platforms like Valo Motion \cite{valomotion2025} have successfully integrated vision-based mixed reality exergames into public spaces. Their systems, which gamify activities such as wall climbing and trampoline jumping, demonstrate the market viability of digitally augmented exercise in gyms, family entertainment centers, and activity parks, a potential future deployment avenue for our research. Notably, Valo Motion's sales material emphasizes that their attractions are unattended and require no headsets or controllers to operate, offering both low operational costs and easy entry for players \cite{valomotion2025fec}. This motivates our focus on a screen-based and vision-based setup instead of VR hardware.

%% file: 3_Design.tex
\section{Design}

This section describes our system and game design. We start by discussing our design goals, followed by an overview of the design process, challenges, technical setup, and motivation for using a vision-based approach. We conclude by detailing the design of various movement mechanics in the game.

\subsection{Design Goals}

Our work is motivated by two primary goals. First, we aim to develop a human-quadruped locomotion system that bridges the morphological gap between bipedal players and four-legged avatars. By utilizing a supine, CV-based configuration, we seek to provide a high-fidelity interface that avoids the constraints of traditional methods, allowing full 4-limb control. Second, we aim to develop an exergame that uses this system to explore new ways to exercise one's core muscles, which have received relatively little attention in exergames.

Our goal is not to compare against a baseline or to prove that our approach is superior in terms of any metric. Rather, in the Research through Design (RtD) tradition \cite{krogh2015rtd}, our goal is to probe and better understand the unexplored design space through the prototype that we developed.

\subsection{Design Overview} \label{designOverview}

Our design stages can be explained using the Double Diamond design framework \cite{designcouncil2005double}: 

\begin{enumerate}
        \item \textbf{Stage 1 - Discovery:} We began by investigating the limitations of traditional locomotion interfaces, observing human supine motion patterns, and analyzing quadrupedal gait dynamics. This step helped to identify the specific biomechanical challenges involved in effectively mapping human motion to quadrupedal movement.
        
        \item \textbf{Stage 2 - Definition:} We established the primary goal of enabling a supine user to control a quadruped avatar across states such as walking, galloping, and jumping. We defined critical constraints, such as ensuring limb freedom and safety, and selected key metrics, including responsiveness and immersion, to guide the system design.
        
        \item \textbf{Stage 3 - Development:} We developed a motion capture system that tracked human motion input. We then built and iteratively refined human to quadruped motion translation algorithms for movement inference and procedural limb mapping across 2D and 3D virtual environments.
        
        \item \textbf{Stage 4 - Delivery:} To avoid tracking glitches inherent in full 3D movement (with strafing and rotation around the vertical axis) caused by movement tracking imperfections, we finalized a 2D side-scrolling platformer design (forward movement and jumping) for a more fun and robust user experience. We also conducted a user study to evaluate the strengths, limitations, and future development potential of the final prototype.
\end{enumerate}

\subsection{Design Challenges}

Mapping non-standing humanoid motion onto quadrupedal characters presents two core challenges. Firstly, accurately inferring a target quadruped’s velocity, posture, and gait from raw humanoid motion data is non-trivial due to the fundamental differences in limb configuration, center of mass, and movement dynamics between bipeds and quadrupeds \cite{Gleicher_retarget_1998, Alexander_AnimalLocomotion_2003, yamane_synthesize_2004}. 

Secondly, ensuring that the approach generalizes across diverse users with varying body proportions and movement patterns while maintaining real-time responsiveness and stability often requires advanced calibration, adaptive retargeting, or possibly learning-based models that can adapt to individual behavioral nuances \cite{egan_dogcode_2024, luo_CARL_2020, holden_synthesis_2023}.


\subsection{Physical and Technical Setup}

\begin{figure}
  \centering
  \includegraphics[width = 0.7 \linewidth]{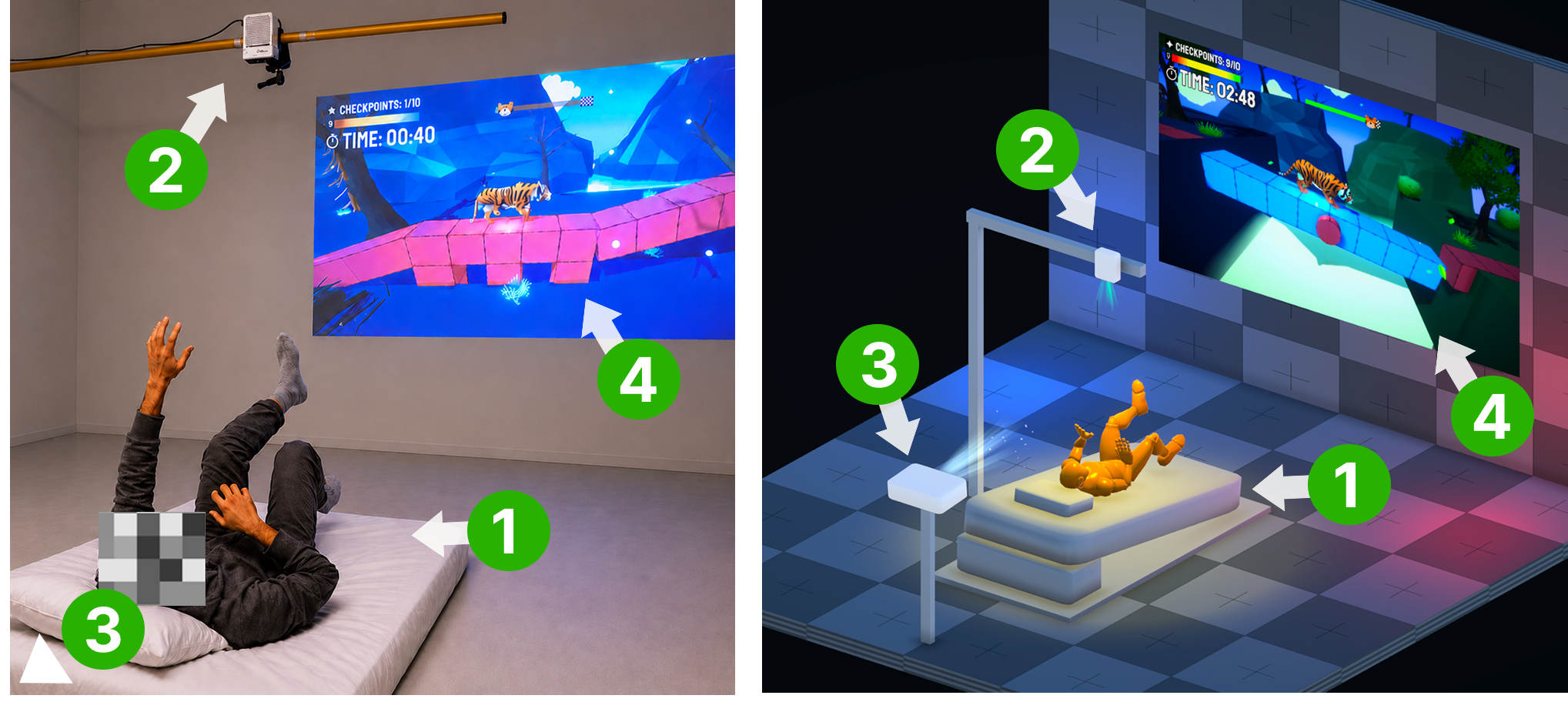}
  \caption{User Study Setup Layout (Left: Real setup, Right: 3D isometric visualization for clarity). 1. User performing movement in supine position over foam mattress for comfort, 2. Depth sensor positioned above the user, aimed downwards, 3. Video projector for displaying game output, 4. Projection wall facing the user.}
  \Description{Test Setup Layout.}
  \label{fig:currentSetup}
\end{figure}

The physical design, illustrated in \Cref{fig:currentSetup}, did not change substantially during the process. To position the sensor far enough over the player to provide sufficient field of view, it was mounted on a broomstick, which was in turn mounted on a large tripod with sandbags to hold the base steady. For the user's comfort and to provide a reasonably comfortable viewing angle of the game, we used two foam mattresses and a pillow. The game graphics were projected onto the screen.

The hardware used to develop the project includes a desktop PC (GPU: NVIDIA GeForce RTX 2060 SUPER, CPU: Intel Xeon W-2133 3.60GHz, RAM: 32GB), and an Azure Kinect DK depth camera developer kit. The game was developed using the Unity game engine (version 2022.3.42f1) by Unity Technologies \cite{unity}.

\subsection{Utilizing Computer Vision and Depth Estimation}

To prioritize accessibility and avoid restrictive marker-based setups, we adopted a computer vision approach using the Azure Kinect DK RGB-D camera \cite{han_cvkinectreview_2013, AzureKinectDK, microsoft_azure_kinect_bodytrack_2024}. The Azure Kinect provides superior skeletal tracking \cite{albert_kinectperformance_2020}, particularly for supine users \cite{shi_kinectlying_2023}. While the original developer kit has been discontinued in late 2023, its depth-sensing technology remains commercially available through partner devices like the Orbbec Femto Bolt \cite{orbbec2023femtobolt}.

Although the Azure Kinect's depth sensor can track input data in 6 degrees of freedom (6-DOF) for 32 joints, our velocity calculations rely primarily on the four limb end effectors, with an additional eleven joints tracked solely for debugging visualization. We positioned the camera directly above the user to prevent limb occlusion in the camera view and to minimize the installation footprint for commercial deployment. Moreover, the top mounted camera setup easily supports a multiplayer future extension with two or more players side-by-side, which is impossible with a side camera due to players occluding each other.

\subsection{Movement Mechanics}

Our primary technical challenge is tracking and translating the user’s limb movements into corresponding quadruped avatar motion. Our movement mapping approach can be summarized as: 
\begin{enumerate}[leftmargin=*]
\item Movement emerges from physics simulation that also conveniently handles gravity and collisions. 

\item We use Inverse Kinematics (IK) to map the tracked user feet and hand positions to a target pose for the quadrupedal character. This pose is used as joint angle targets for the physics simulated skeleton's joint actuators. 

\item To optimize responsiveness and game feel, we adjust the physics simulation in two ways:
\begin{itemize}
    \item Instead of relying on physics simulated friction, which we have found to cause constant slipping of the feet, we implement our own "pseudo friction" where foot movement pushes the character forward if it is close enough to the ground.
    \item For immediate responsiveness, we instantly override the avatar's velocity for jumps and forward movement instead of applying control forces that gradually accelerate towards desired velocity.
\end{itemize}
\end{enumerate}

Below, we provide additional details about how forward locomotion and jumping are implemented.

\begin{figure}
  \centering
  \includegraphics[width=1\linewidth]{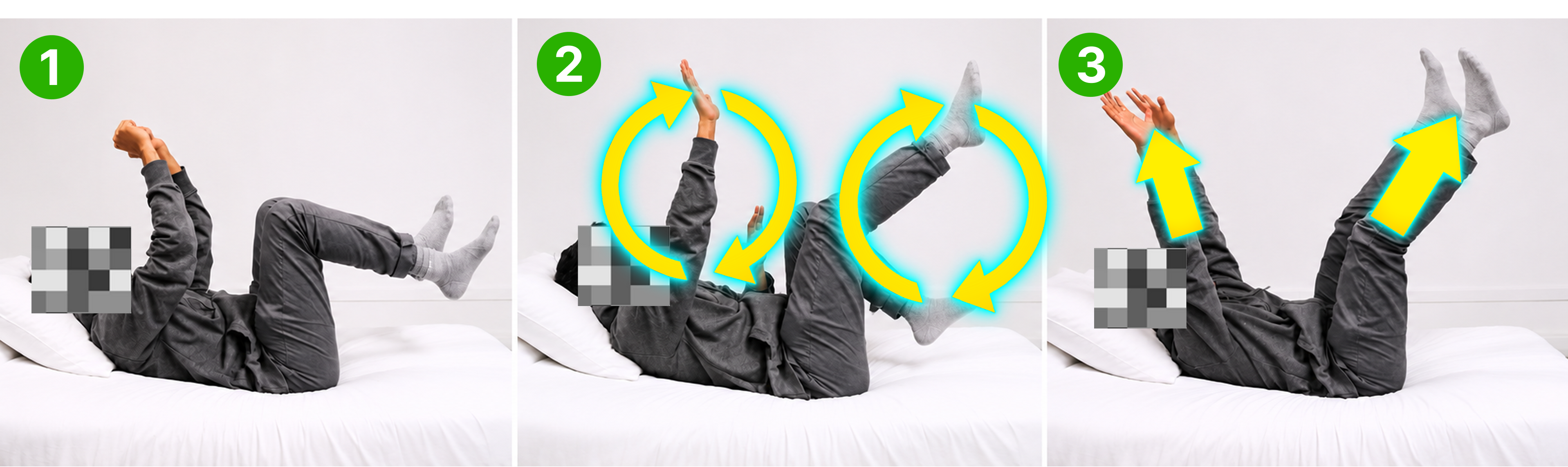}
  \caption{Player movement. 1) Calibration pose held at the beginning of the game for 3 seconds, 2) Forward cyclic locomotion with all four limbs, 3) Movement for upward jump with all four limbs. Arrows have been added to clarify cyclic motion and jump motion directions.}
  \label{fig:intendedMovement}
  \Description{}
\end{figure} 

\subsubsection{Forward/backward locomotion} \label{forward}

The user drives forward motion naturally, with the speed of limb movement directly controlling the avatar's velocity. The algorithm calculates a weight $w$ for each limb to determine how much influence it should have on overall movement as:

\begin{equation}
    w = \max\left(0, 1 - \frac{d}{c}\right) \label{eq:clamped_weight},
\end{equation}

where $d$ is the limb's distance from the ground and $c$ is the ground contact zone thickness (0.25m in our case). If a limb is in touch with the ground ($d=0$), w is 1.0 (full influence). If a limb is too far from the ground ($d>=c$), w is 0.0 (no influence). The limbs closest to the ground have the most influence.

The target velocity to apply to the avatar equals a weighted average of all limb velocities:

\begin{equation}
    v_{target} = b_{xz}\frac{\sum_i w_i v_i}{\sum_i w_i},
\end{equation}

where $v$ denotes the tracked limb velocity, $i$ is the limb index and $b_{xz}$ is a velocity boost multiplier. Additionally, to allow the physics-simulated friction to slow the avatar down instead of it stopping immediately when the user stops, we set a limb's weight to zero if it is not moving faster than a threshold value. If the sum of all weights is zero, the physics simulation velocity is not overridden. 



\subsubsection{Jumping}
Our jump logic can be summarized as: If the user's limbs are moving fast enough upwards (towards the Azure Kinect) and if the avatar is grounded, we override the avatar's velocity with a jump velocity. Otherwise, the velocity is not touched by the avatar movement code, letting the physics simulation control it according to gravity and collisions. To avoid false negatives, the grounded check includes "coyote time", i.e., it returns true if the avatar is touching the ground or has touched the ground during the last 200ms.

The jump velocity's horizontal and vertical components, $v_{y\_jump}$ and $v_{z\_jump}$, are determined as follows:
\begin{equation}
    \begin{aligned} \label{eq:jump}
        v_{y\_jump} &= clip(b_{y} v_y, v_{y\_avatar}, v_{y\_max}) \\
        v_{z\_jump} &= clip(v_z + b_{z} v_y, v_{z\_avatar}, v_{z\_max}) \\
    \end{aligned}
\end{equation}

where the $clip(x, x_{min}, x_{max})$ function clips $x$ in the range  $[x_{min}, x_{max}]$, $v_{y}$ denotes the vertical tracked velocity of the user's limb (the jump velocity update is performed for each limb, letting players decide to jump using only their hands, legs, or by using all four limbs), $v_{y\_avatar},v_{z\_avatar}$ are the avatar's current vertical and horizontal velocity, $b_y,b_z$ are movement boost multipliers, and $v_{y\_max},v_{z\_max}$ are the maximum allowed avatar vertical and horizontal velocities. 

The $clip$ function ensures that the jump neither slows the avatar down nor exceeds maximum speed limits ($v_{max}$) due to tracking noise. Finally, the $b_z$ multiplier adds a forward lunge to the vertical motion, mimicking a pouncing tiger to help players clear gaps more easily.

%% file: 4_UserStudy.tex
\section{User Study}

Our user study utilized a mixed-methods approach \cite{creswell1999mixedmethod}, integrating both quantitative and qualitative data forms to measure the usability of our proposed locomotion system and to better understand the overall user experience. Consistent with our Research through Design approach, the study is an exploratory one. Our study aimed to answer the three research questions mentioned in \Cref{introduction}. 


\subsection{Study Design}

After signing the informed consent form, each participant encountered a progression of three game levels. It should be noted that the levels did not correspond to different experimental conditions of a within-subjects study; therefore, the order of the levels was not counterbalanced. Data analysis centered on the qualitative thematic analysis of think-aloud data captured during gameplay and responses from a semi-structured interview conducted post-gameplay based on the questions stated in Table \ref{tab:interviewQuestions}. To complement the qualitative findings, we collected quantitative measures, specifically self-reported exercise intensity and responses to the System Usability Scale (SUS), which were used to derive an overall quantitative measure of usability \cite{brooke1996sus}.


The locomotion task involved navigating three game levels consisting of multiple checkpoints, and participants had to use the two primary movements: forward locomotion and jumping. The desired locomotion movements were instructed as illustrated in Figure \ref{fig:intendedMovement}, using short video clips to demonstrate how moving and jumping can be performed.

Participants traversed three untimed levels (see Figure \ref{fig:levelLayout}) built as progressive obstacle courses. Based on pilot testing, we opted for open, low-penalty designs to encourage continuous movement. The stages advanced from basic walking to complex jumping and timing challenges. We maintained a fixed level order rather than counterbalancing, as the sequence was designed to act as an extended tutorial, ensuring all users could master the baseline physical mapping before engaging with advanced platforming challenges.

\subsection{Participants}

Fifteen participants (four female, eleven male) were recruited by reaching out to different communication channels and social circles throughout the university. Each participant was offered a 20 euro gift card to the on-campus university merchandise shop for their contribution.

Thirteen participants were in the age group of 25-34 years, while two belonged to the age group of 18-24 years. The majority of the participants identified themselves as university students, while others identified as university researchers, game designers, or engineers. Three participants reported exercising daily; seven exercise two or three times a week, and five reported that they had not exercised in the past month or so. We did not specify what we exactly meant by "physical exercises"; hence, users were free to interpret it as going to the gym, walking, or running, etc. 

Participants reported diverse prior engagement with exergames that involved some degree of physical movement (mostly with their hands). 
Five participants mentioned playing VR games such as Beat Saber \cite{beatsaber}; three mentioned Wii Sports \cite{nintendo_wiisports_2006} and Wii Fit \cite{nintendo_wiifit_2007}; two mentioned Switch Ring Fit Adventure \cite{RingFitAdventure}, and others mentioned Kinect games such as Dance Dance Revolution \cite{DanceDanceRevolution} and Kinect Sports \cite{KinectSports}. 

\subsection{Data Collection}

The following data was collected during and after the gameplay sessions:

\subsubsection{During Gameplay:}

We used the Borg CR10 perceived exertion scale \cite{borg1990psychophysical} to measure participants' exertion/exhaustion immediately after completing each of the three levels. Participants were asked to report how exhausted they felt on a scale of 0 to 10. The study instructor explained the meanings of the minimum and maximum values on the scale and what exactly was being assessed. 

Participants were also encouraged to think aloud during the gameplay and to comment on moment-to-moment interactions and events. Additionally, we captured gameplay video of the virtual world alongside the user’s voice recordings during gameplay. The voice recordings were transcribed, and the video was analyzed to determine whether the transcripts needed further context. Furthermore, the study instructor took notes about observations, such as different movement styles, how the user approached a particular obstacle, and what they seemed to struggle with. We also collected a range of task performance data during gameplay. This included timestamps for each checkpoint reached, the total number of respawns per level, and the completion time for each level.

\subsubsection{Post-Session Assessment (After all three levels completed):}

Upon completing the entire gameplay session, the following data collection sequence was administered:

\begin{enumerate}
    \item A semi-structured interview was conducted to gather in-depth feedback on the overall experience of the locomotion system. We designed and used the questionnaire in \Cref{tab:interviewQuestions} for the interview.
    \item A brief final survey was administered immediately following the interview, which included the System Usability Scale (SUS) scores and general demographic information. We also requested the users to describe the game movement using three adjectives.
\end{enumerate}

All data was stored anonymously. The voice recordings were deleted after transcription. Participants were identified only by numeric IDs, with no record kept that allows mapping an ID to a name. Only the gameplay video was recorded, and no real world video was captured.

\subsection{Safety and Wellbeing Considerations}

For safety reasons, the recruitment was targeted towards physically active participants who would be comfortable exercising. We advertised the user study as a "core workout exergame". To ensure participant safety and wellbeing during the rigorous physical activity, a study facilitator actively monitored users for overexertion, with explicit instructions that they could rest or withdraw at any time. Furthermore, the system's supine posture inherently mitigates common exergaming hazards, effectively eliminating the risks of losing balance, falling, or colliding with real-world objects.

\subsection{Methods}

\subsubsection{Qualitative Analysis:}

We analyzed the aforementioned gameplay discussions and the post gameplay semi-structured interviews using Reflexive Thematic Analysis (RTA) \cite{braun2006using, braun2019reflecting, byrne2022workedRTA}, following the standard progression of data familiarization, data coding, theme generation, reviewing and refining the themes, and reporting with illustrative quotes \cite{braun2006using}. The transcripts were analyzed using inductive coding, without a predefined codebook. Reflexive TA was selected over other approaches, such as codebook TA or coding reliability TA, because it provides the interpretive flexibility necessary for deep contextual analysis \cite{braun2019reflecting}. Following standard RTA guidance, our data coding and analysis were performed by a single author to ensure a deeply reflexive, continuous, and unified interpretation of the qualitative data \cite{braun2019reflecting}. 

\subsubsection{Quantitative Analysis:}

Quantitative data gathered from the survey responses, Borg CR10 scores, SUS scores, and performance metrics were analyzed using descriptive statistics, specifically averages, means, medians, and standard deviations. Due to our limited sample size (N=15), we do not perform any statistical significance tests. We also do not perform any comparisons with a baseline or existing locomotion systems since our goal is not to claim that our approach is superior to previous work, but rather to explore and understand the novel design space of quadrupedal supine exergames. Therefore, this approach allows us to clearly describe the observed patterns in player experience, usability, and control, prompting hypotheses and research questions for future work rather than drawing conclusions. 

%% file: 5_Results.tex
\section{Results}

\subsection{Qualitative Results}

Qualitative analysis provided deep insights into the player experience, capturing reflections regarding control, exertion, embodiment, immersion, and ergonomics. Through a reflexive thematic analysis, we constructed 287 unique codes from 546 individual quotations.

To accurately capture the nuances of the player experience, we iteratively synthesized these codes into overarching qualitative themes that tell the story of the player's embodied journey. Table \ref{tab:themes_summary} summarizes these core themes, which encapsulate the interplay between physical exertion, avatar control, usability challenges, mastery, and player suggestions, along with representative participant quotes. When asked to describe the overall in-game movement with three adjectives, participant responses clustered around terms like "fun," "engaging," and "novel," alongside strenuous attributes such as "exhausting" and "demanding." These frequency results can be visually summarized as the word cloud in Figure \ref{fig:wordcloud}.

The following subsections (\ref{sec:rq1}, \ref{sec:rq2}, and \ref{sec:rq3}) unpack these themes in detail, associating them with each of our core research questions \ref{introduction}.

\subsection{RQ1: Player Observations and the Overall Locomotion Experience}\label{sec:rq1}

\subsubsection{Player-Avatar Resonance and Dissonance}
Movement was often described as smooth, responsive, intuitive, and providing a feeling of control. Participants enjoyed the feeling of a direct, precise mapping where the character's actions mirrored their physical intent (\textit{"It feels quite good. When I want to jump, it jumps. When I want to run forward, it runs forward" [P14]}). This synchronization often felt natural and met user expectations, with some participants specifically highlighting the lack of perceived latency helping to maintain a deep connection to the avatar (\textit{"I felt it's, uh, kinda very well connected between the movement and the translation on the tiger" [P6]}). 

However, several participants indicated that the tiger movement felt unnatural or biologically inconsistent (\textit{"Because sometimes it feels a little bit off. It doesn't really follow like the tiger anatomical... So, it doesn't look natural." [P6]}), leading one participant to feel a separation of agency (\textit{"It feels like he's trying to emulate me, but it's not that I'm controlling the tiger" [P9]}). Tracking issues also impacted the experience, with some participants reporting a sense of input lag, particularly during jumps. These delays could lead to a momentary loss of control and break the user's immersion (\textit{"Whenever there is like, input delay then I can snap out of it" [P1]}), resulting in feelings of intermittent control.

Finally, physical ergonomics played a role in this dissonance. While most found the supine position natural and sometimes more comfortable than a traditional VR setup (\textit{"I think this is definitely easier and more comfortable than a VR setup" [P7]}), users noted issues with physically slipping from their default position. This need for constant repositioning resulted in unintentional inputs and reminded players of their physical constraints.

\subsubsection{From Friction to Flow: Navigating Mastery}
The jump mechanic was identified as the system's most crucial and polarizing element. For some participants, jumping was initially difficult, feeling delayed or hard to estimate in direction. We identified issues stemming from: (1) the jump was not properly tracked, (2) user did not execute the intended jump motion, and (3) user jumped too late and missed the ground contact required to initiate the jump. Notably, during pilot testing, some participants returned their limbs to the "ready" position too late, pushing the avatar backwards and causing frustration. 

Yet, overcoming this initial friction was highly rewarding. Once players learned to jump properly, it was considered responsive, fun, and satisfying (\textit{"Um, it did feel very satisfying to jump. Especially if I was, like, able to get the motion right and I jumped really far, it felt really good" [P10]}). Participants derived a strong sense of empowerment from the system’s exaggerated mechanics (\textit{"And then when I jumped, it clicked. Due to excitement, I would be doing more and then jumping super high. That was my favourite moment" [P4]}). Mastering the movement to achieve finer control such as navigating tougher obstacles or skipping sections transformed physical effort into rewarding in-game moments. This progression from "easy to start but hard to master" ultimately led to a sustained state of flow (\textit{"...because it felt like I was in the flow which also helped me keep going" [P15]}).

\subsection{RQ2: The Locomotion Technique as a Core Muscle Workout}\label{sec:rq2}

\subsubsection{Core Workout Effectiveness}
Participants frequently reported that the game provided a demanding core workout, effectively targeting the abdominals, thighs, and pectorals. The physical intensity of the study was apparent to almost all participants, who recognized the system as a genuine physical challenge. As one user stated, \textit{"When I was doing the movement on the ground, it feels like I was doing real exercise" [P6]}. Some users even speculated on the long-term health benefits of the system, such as weight management (\textit{"I feel like this is a very good one to reduce belly fat" [P11]}).

\subsubsection{The "Stealth Exercise" Illusion}
Crucially, the gamified environment provided the necessary motivation to sustain this rigorous activity, which many noted would be difficult to endure in a traditional exercise context (\textit{"I would find very hard to motivate myself doing that kind of movement without that gaming element" [P8]}). Participants generally appreciated this "stealth exercise" nature of the game, where the intense focus required for platforming gameplay successfully masked their fatigue. As one participant emphasized, \textit{"It gets you exercising in a way that doesn't feel like exercise. You were asking me if I wanted to take breaks but I just said, 'No, no, let's keep going'" [P14]}.

\subsubsection{Physical Barriers \& Acute Fatigue}
Despite the motivational benefits of the game, physical condition was cited as a firm boundary for some participants. Feedback included comments about the movement being thoroughly exhausting (\textit{"Uh, after the tutorial level, uh, I was a bit exhausted. And after the game ends, I was really exhausted" [P1]}). For others, the exertion caused discomfort due to pre-existing, long-term physical conditions (\textit{"Um, for me, it got like painful pretty quickly. I was primarily stopped by like having lower back pain" [P9]}). Furthermore, the novelty of the quadrupedal movement resulted in acute physical sensations, such as unexpected leg fatigue (\textit{"Naturally, legs were getting tired. I didn't expect my legs to get that tired" [P15]}).

\subsection{RQ3: Future Improvements and System Expansions}\label{sec:rq3}

\subsubsection{Yearning for Expressive Freedom}
Participant feedback highlighted a clear demand for expanded locomotion capabilities, specifically requests for backward, lateral, and climbing mechanics to enhance navigational freedom. Users suggested that lateral movement could improve the experience (\textit{"I think I would like to have more lateral movement to be able to go left and right more freely" [P13]}), while others envisioned vertical actions like climbing (\textit{"maybe climbing up a tree or a cliff something that a tiger would naturally do" [P6]}). Adding mid-air control was also mentioned to improve the platforming feel.

Beyond locomotion, participants requested "tiger-specific" gameplay activities to better leverage the animal embodiment, such as chasing or interacting with the environment (\textit{"add more tiger like activities... like hunting some deer... drinking water to regain stamina" [P2]}). Players also desired more levels for longer playtime and the addition of enhanced audio or haptic feedback. To mitigate tracking issues caused by user movement, participants suggested mid-game recalibration features (\textit{"if I'm slipping from the neutral position, then maybe it could recalibrate mid-game" [P10]}). Finally, a strong desire for social engagement emerged; players requested multiplayer game modes and competitive structures to drive long-term engagement (\textit{"I wanted to see how I did compared to what other players have done. So I would want to see all time high scores" [P15]}).

\subsection{Quantitative Results}

\subsubsection{System Usability Scale (SUS) Score}

The overall SUS score across all participants (N=15) is 77.33 (SD = 12.66).
Based on the Sauro and Lewis curved grading scale \cite{lewis2009factor}, this score corresponds to a school grade of B+ (Good), percentile range 80–84\%, meaning the system is perceived as more usable than approximately 80–84\% of systems tested.

\begin{figure}
  \centering
  \includegraphics[width=0.8\linewidth]{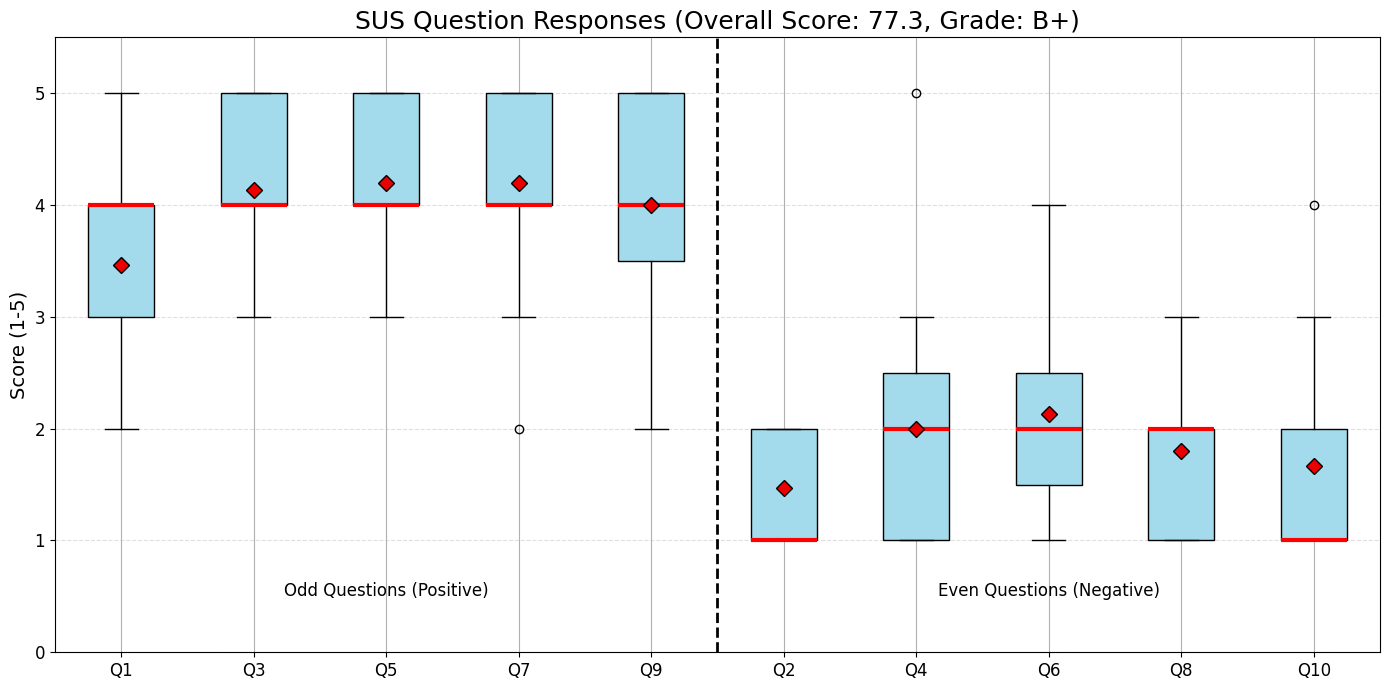}
  \caption{Box plot visualizing the distribution of responses for each SUS question (Q1–Q10). Red lines represent the median response, red diamonds represent the mean response, boxes show the interquartile range and circles represent the outliers. For the odd items on the left (Q1, Q3, Q5, Q7, Q9), higher is better. For the even items on the right (Q2, Q4, Q6, Q8, Q10), lower is better.}
  \Description{}
  \label{fig:sus}
\end{figure}

Figure \ref{fig:sus} visualizes the distribution of responses for each SUS question (Q1–Q10) \cite{lewis2018SUS}, where odd-numbered questions (Q1, Q3, etc.) are positive (higher is better), and even-numbered questions (Q2, Q4, etc.) are negative (lower is better).
    
\subsubsection{Borg CR10}

Figure \ref{fig:rpe} illustrates the distribution of Borg CR10 scores reported at the end of each level. The results suggest that our prototype indeed provides an intensive core workout, with exhaustion growing over the levels. For level one, the average Borg CR10 score was 3.80 (Median = 3.40, SD = 2.83, Min = 0, Max = 10). For level two, the average was 6.47 (Median = 7.00, SD = 1.93, Min = 3, Max = 9). Finally, for level three, the average score was 7.26 (Median = 7.6, SD = 1.77, Min = 4, Max = 10). 
Note that the maximum score of 10 recorded for level one (tutorial) represents one participant's initial difficulty in executing a jump. The participant's exertion score subsequently decreased as they became proficient with the jump mechanic over the next levels. 

\subsubsection{Level Completion Time}

All fifteen participants successfully passed all the checkpoints across three levels, completing the full game. The average completion time for level two was 128 seconds (Median = 86, SD = 109, Min = 64, Max = 485). The average completion time for level three was 138 seconds (Median = 128, SD = 64, Min = 73, Max = 322). We considered level one to be the tutorial level and did not keep track of the completion times for it. There was no maximum time limit for any of the levels.

%% file: 6_Discussion.tex
\section{Discussion} \label{discussion}

Considering both the qualitative and quantitative results, we can now summarize our findings to address our core research questions \ref{introduction}, contextualizing the player experience within existing HCI frameworks and outlining possible directions for future research. 

\subsection{Embodiment: Agency vs. Body Ownership}

A recurring theme in our qualitative data regarding the locomotion experience (RQ1) was the division between players feeling in control of the avatar versus feeling like an animal. We contextualize this using \citet{kilteni2012sense}'s framework for the Sense of Embodiment (SoE), which divides embodiment into three sub-components: the sense of agency (the feeling of motor control), the sense of body ownership (the feeling that the virtual body is one's own), and the sense of self-location.

Our continuous 1:1 motion mapping provided players with a high sense of agency. Participants generally felt their physical intent immediately and accurately translated to the tiger's movement. However, the sense of body ownership was occasionally hindered by the anatomical mismatch between human and feline biomechanics. Consistent with \citet{Krekhov_VRAnimals_2018}'s findings on non-human avatars, we observed that while exaggerated physics empower players, true body ownership remains difficult to achieve when proprioceptive feedback misaligns with visual feedback \cite{gorisse2017first}.

Despite occasional breaks in embodiment, the exaggerated mechanics ultimately served as a powerful gameplay driver. As players transitioned from the initial friction of learning the jump mechanic to a state of mastery, they entered a satisfying state of flow. This indicates that in full-body exergames, absolute biological realism may be less critical to player enjoyment than providing a responsive, empowering, and highly exaggerated physics system that rewards physical execution.

\subsection{Gamifying Exertion: The ``Stealth Exercise'' Effect}

Addressing our second research question (RQ2), our results indicate that the system successfully transformed a rigorous core workout into a playful experience. Participants reached high exertion levels (Borg CR10 scores escalating up to 10) yet exhibited high motivation to continue playing. This aligns seamlessly with the dual-flow model of exergaming \cite{sinclair2007considerations}, where cognitive immersion in platforming challenges effectively distracts users from acute physical fatigue.

Extending \citet{reetu_abs_2023}, our work demonstrates that gamified physical-to-virtual mappings can drive intense, sustained athletic engagement without immersive hardware or wearable trackers. This reinforces foundational exertion game principles \cite{mueller2011designing}, specifically the concept that tightly coupling strenuous bodily movement with immediate virtual feedback successfully transforms physical strain into rewarding play. However, designers must remain conscious of the physiological limits of this illusion as our qualitative findings indicate, the "stealth exercise" effect can be abruptly shattered by acute muscle fatigue or pre-existing physical conditions, highlighting the need for scalable difficulty in future iterations.

\subsection{Design Implications: Ergonomics and Expressive Freedom}

Looking toward future iterations (RQ3), player feedback highlighted two main avenues for improvement: hardware ergonomics and software expressiveness.

On the hardware front, our screen-based approach was strongly validated by user feedback, circumventing the severe visual-vestibular sensory conflicts often induced by playing VR games in a supine position \cite{luo_sensoryconflict_2022}. However, translating fast-paced exergames to unconventional floor-based postures requires specific ergonomic scaffolding. Because visual targets naturally guide human limb positioning, sub-optimal screen placement can strain proprioception \cite{han2016proprioception} and cause players to physically drift out of the sensor's tracking frame. 

\begin{figure}
  \centering
  \includegraphics[width=0.5\linewidth]{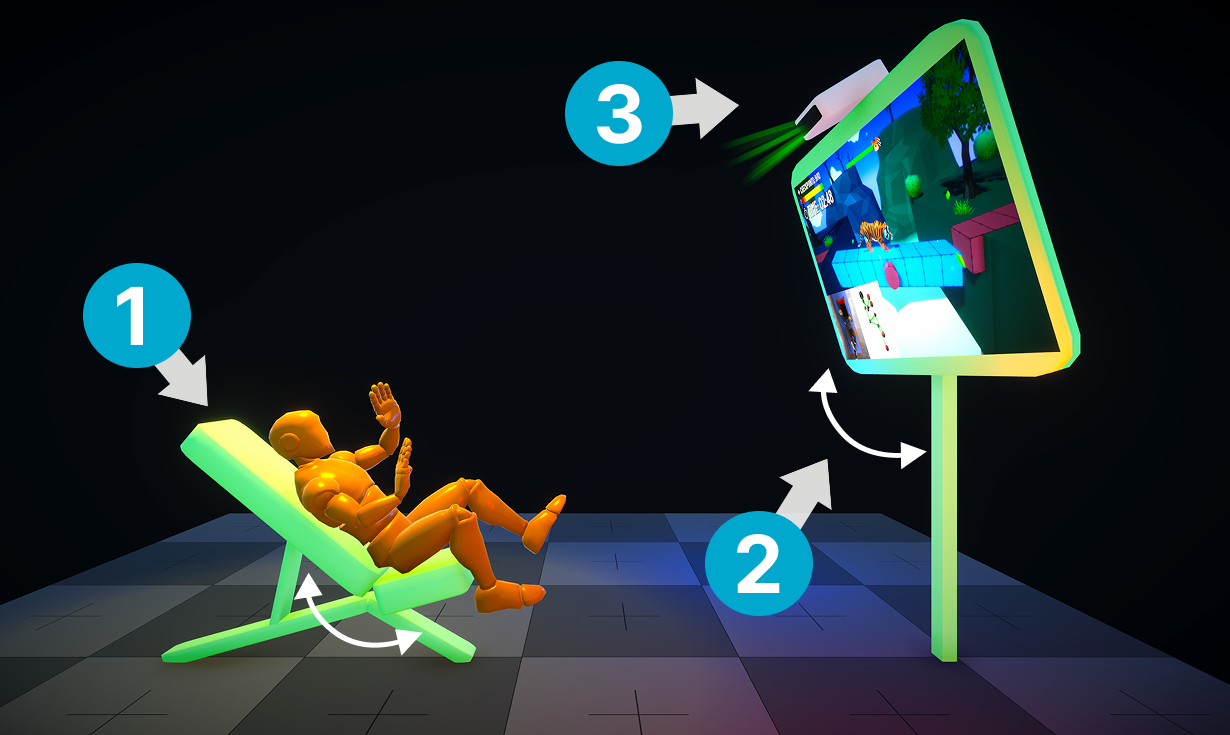}
  \caption{Visualization of a hypothesized revised hardware setup consisting of 1) an adjustable recliner, 2) a mount that allows tilting the screen to face the player, and 3) the depth sensor mounted on the screen.}
  \Description{}
  \label{fig:revisedSetup}
\end{figure}

Future work must prioritize the design of the physical setup. \Cref{fig:revisedSetup} shows a hypothetical revised setup that could enhance the overall usability of our system. We propose an adjustable recliner that is higher than the mattress and easier to get into and out of. This enables players to scale their exertion (a more upright posture reduces leg-movement effort) and mitigates the physical slippage identified in our current setup. We also propose a tilted screen that doubles as a mount for the depth sensor instead of requiring a separate stand.

On the software front, players expressed a strong yearning for expressive freedom to fully realize their animal embodiment. Future systems should expand navigational mechanics (e.g., lateral strafing, climbing) to give players a wider repertoire of spatial problem-solving tools. Furthermore, integrating competitive social elements (multiplayer modes, high scores) will be critical for deepening immersion and driving long-term player retention.

%% file: 7_Limitations_and_Future_Work.tex
\section{Limitations and Future Work}

While our sample of 15 participants aligns with standard qualitative HCI guidelines \cite{caine2016local} and is sufficient for usability evaluation \cite{hwang2010number}, the quantitative data we have gathered are used for descriptive purposes, as a larger sample is required for conducting any comparative analyses or statistical evaluations. Player adaptation remains a critical factor influencing system performance. Individual differences such as body size and motor behavior can result in unconventional movement patterns that challenge the system’s movement detection and mapping accuracy. Furthermore, our approach is constrained by the Azure Kinect depth sensor, which suffers from higher latency, lower spatial resolution, and greater occlusion susceptibility compared to dedicated wearable trackers, which can reduce tracking fidelity and responsiveness during rapid or complex movements. Expanding the system to fully explorative 3D environments or deploying it commercially will necessitate a custom, optimized tracking solution capable of capturing rapid, complex movements in order to maximize accuracy and minimize latency and glitches. Finally, although the Borg CR10 scale \cite{borg1990psychophysical} is a validated metric for perceived exertion in HCI research, future studies should incorporate objective physiological measures, such as heart rate monitoring, to more robustly quantify exercise intensity.

Future work should focus on three primary areas to address current limitations and enhance adaptability, precision, and accessibility. First, tracking latency and noise can be mitigated by developing a custom computer vision solution optimized for hand and foot positioning, bypassing the need for full-body skeletal tracking. Second, user intent and gait state inference could be improved via machine learning. To circumvent labor-intensive dataset labeling, data collection could be gamified, prompting users to follow rhythm-game-style choreography to automatically sync physical movements with intended inputs. Finally, participants expressed a strong desire for social interaction, suggesting that one could enhance motivation and enjoyment by introducing competitive or collaborative gameplay. The system could easily expand to support side-by-side multiplayer gameplay tracked simultaneously by a single depth sensor.

%% file: 8_Conclusion.tex
\section{Conclusion}

In this paper, we presented a novel human-to-quadruped locomotion system and an associated exergame designed to provide an intensive core workout. By utilizing a computer vision-based approach, our system circumvents the need for wearable sensors and VR headsets, successfully avoiding the visual-vestibular sensory conflicts common in immersive interaction systems.

Our mixed-methods user study demonstrated that tightly coupling full-body physical movement to quadrupedal avatar control creates powerful gamified exertion. Participants reached high exertion levels while maintaining high engagement, driven by the cognitive distraction of the gameplay and an empowering sense of motor agency. While true non-human body ownership remained limited by inherent anatomical constraints, responsive control mapping proved sufficient to deliver a satisfying and physically demanding player experience.

Beyond our specific implementation, this research contributes broader insights for the design of future exergames and embodied interactions. We highlight the necessity of prioritizing continuous, predictable motor agency over strict anatomical realism when mapping bipedal human movements to non-human avatars. Future work should consider exploring refined computer vision techniques to accommodate highly diverse physical movement patterns and investigating how the integration of social multiplayer dynamics might further improve the locomotion system.


%% file: Appendix.tex
\section{Theme Summary}

\begin{table*}[h!]
\centering
\caption{Summary of generated reflexive themes capturing the overarching player experience, along with representative quotes.}
\label{tab:themes_summary}
\renewcommand{\arraystretch}{1.4}
\begin{tabular}{@{} p{3.5cm} p{5.5cm} p{5.5cm} @{}}
\toprule
\textbf{Theme} & \textbf{Summary} & \textbf{Quotes} \\
\midrule

The "Stealth Exercise" Illusion & The immersive, gamified environment acted as a cognitive distraction, transforming a grueling physical core workout into a rewarding and highly motivating play experience. & \textit{``I think it was quite unique and fun because I was also engaging my core, but I had not realized this because I was too focused on the game.''} [P15] \\

Core Workout Effectiveness & Participants recognized the system as a genuine, demanding physical workout that effectively engaged abdominal, thigh, and pectoral muscles. & \textit{``When I was doing the movement on the ground, it feels like I was doing real exercise.''} [P6] \\

From Friction to Flow: Navigating Mastery & The initial physical and technical struggle to execute complex movements (like jumping) eventually gave way to a satisfying state of flow, fueled by empowering, exaggerated virtual physics. & \textit{``And then when I jumped, it clicked. Due to excitement, I would be doing more and then jumping super high. That was my favourite moment.''} [P4] \\

Player-Avatar Resonance and Dissonance & While players were fully immersed in the experience for the most part, some players experienced a fragile sense of embodiment at times. Seamless motor control created deep immersion, but tracking lag, anatomical mismatches, and physical slipping quickly shattered the illusion. & \textit{``It feels like he's trying to emulate me, but it's not that I'm controlling the tiger.''} [P9]

\textit{``It feels quite good. When I want to jump, it jumps. When I want to run forward, it runs forward.''} [P14]

\textit{``I feel like there's an input lag when I jump. Whenever there is like, input delay then I can snap out of it.''} [P1] \\

Physical Barriers \& Acute Fatigue & While the game masked exertion for many, some players reached physical limits, experiencing sudden exhaustion or unexpected muscle fatigue. & \textit{``Naturally, legs were getting tired. I didn't expect my legs to get that tired.''} [P15] \\

Yearning for Expressive Freedom & As players acclimated to the quadrupedal form, they sought richer, species-specific interactions and social elements to more fully realize the fantasy of embodying the animal. & \textit{``Add more tiger like activities like hunting a prey, drinking water to regain stamina.''} [P2]

\textit{``I wanted to see how I did compared to what other players have done. So I would want to see all time high scores.''} [P15] \\

\bottomrule
\end{tabular}
\end{table*}

\clearpage

\section{Movement Adjectives Visualization}

\begin{figure}[!h]
  \centering
  \includegraphics[width=0.6\linewidth]{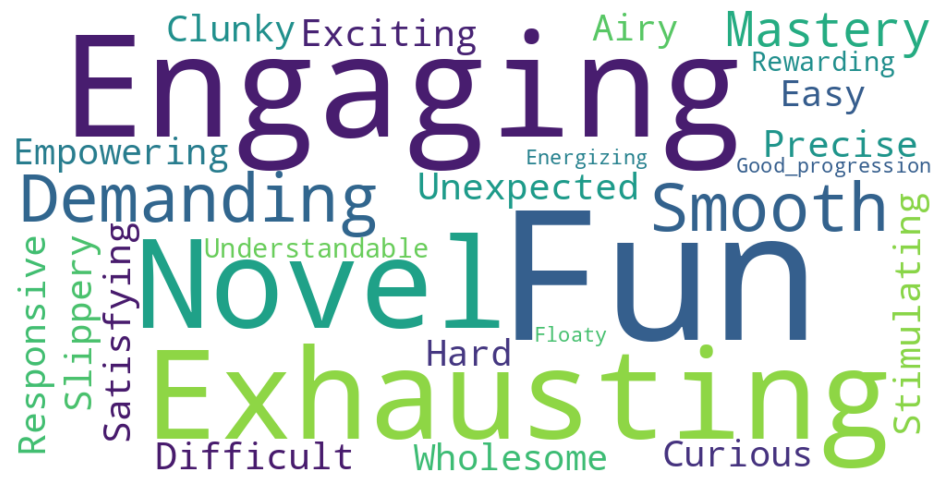}
  \caption{Word cloud visualization of movement adjectives as mentioned by participants. The size of each word is proportional to its frequency of mention in the interviews.}
  \Description{}
  \label{fig:wordcloud}
\end{figure}

\section{Borg CR10 Distribution Visualization}

\begin{figure}[!h]
  \centering
  \includegraphics[width= 0.6\linewidth]{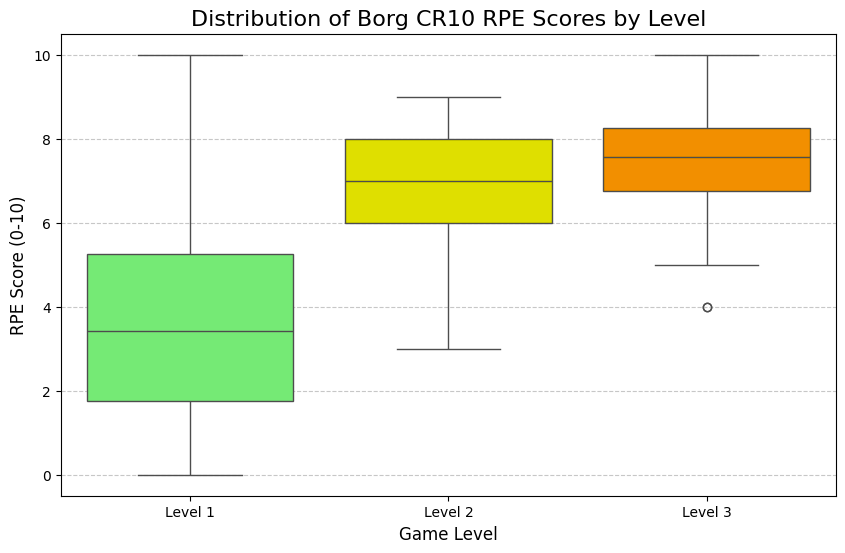}
  \caption{Box plot visualizing distribution of Borg CR10 scores as reported by the participants after each level (level 1, level 2, level 3) where 0 represents no effort and 10 represents maximum exertion. Circle represents the outlier.}
  \Description{Box plot visualizing distribution of Borg CR10 RPE scores as reported by the participants after each level.}
  \label{fig:rpe}
\end{figure}

\clearpage

\section{Level Design}

\begin{figure}[h!]
  \centering
  \includegraphics[width = 0.8 \linewidth]{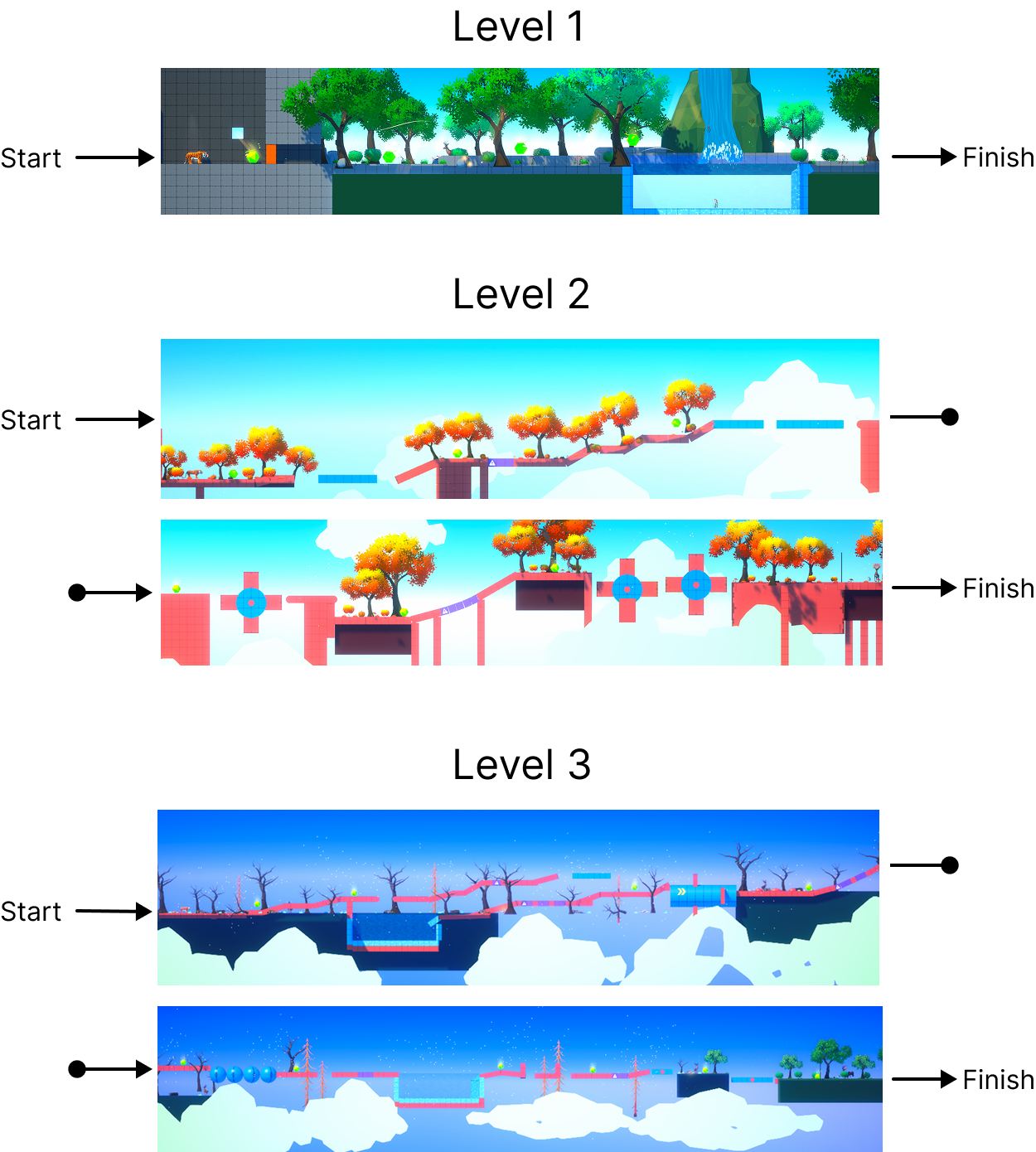}
  \caption{Level 1, level 2 and level 3 design overview from the side highlighting key areas having their unique locomotion challenges.}
  \Description{}
  \label{fig:levelLayout}
\end{figure}

\clearpage

\section{Interview Questions}

\begingroup
\renewcommand{\arraystretch}{2}
\begin{table}[h!]
\centering
\caption{Interview questions asked during the user study.}
\label{tab:interviewQuestions}
\begin{tabular}{|l|p{10cm}|}

\hline

\textbf{Section} & \textbf{Questions}\\ 

\hline

\textbf{Warm-up} & \begin{enumerate}
  \item How often do you engage in physical exercise?
  \item Have you played any exergame before? Which ones?
\end{enumerate} \\[8pt]

\hline

\textbf{User Experience} & \begin{enumerate}
  \item Please describe your overall experience of using the novel locomotion system.
  \item What was the most memorable part for you, if there were any?
  \item What positive aspects did you find about this movement method, if there were any?
  \item What negative aspects did you find about this movement method, if there were any?
  \item What would you like to change about this movement technique, if there is anything?
\end{enumerate} \\[8pt]
\hline

\textbf{Embodiment} & \begin{enumerate}
  \item How did the overall in-game movement feel like? Please describe with at least 3 adjectives.
  \item How did you physically feel when you moved while lying down?
  \item Did you feel in control of the quadruped character?
\end{enumerate} \\[8pt]
\hline


\textbf{Wrap up} & \begin{enumerate}
  \item Thanks, that would be everything for the interview. Do you have any final comments you would like to share?
\end{enumerate} \\[8pt]
\hline

\end{tabular}
\end{table}
\endgroup